\documentclass[12pt]{article}

\ifx\pdfoutput\undefined
\usepackage[dvips,bookmarks]{hyperref}
\else
\usepackage{hyperref}
\fi
\hypersetup{colorlinks=false,bookmarksopen,bookmarksnumbered,citecolor=blue,
   pdfstartview=FitH}

\usepackage[dvips]{graphicx}
\usepackage{latexsym}
\usepackage{amssymb,amsfonts,amsmath}
\usepackage{graphicx} 
\usepackage{indentfirst}

 \usepackage{bbm}

\parskip=\medskipamount

\newcommand {\cD}{{\cal D}}

\newcommand {\cF}{{\cal F}}
\newcommand {\cG}{{\cal G}}

\newcommand {\cL}{{\cal L}}
\newcommand {\cM}{{\cal M}}
\newcommand {\cN}{{\cal N}}

\newcommand {\cV}{{\cal V}}
\newcommand {\cW}{{\cal W}}

\def\a{\alpha}
\def \bi{\bibitem}

\def\b{\beta}

\def\d{\delta}
\def\e{\epsilon}

\def\g{\gamma}
\def\G{\Gamma}

\def\k{\kappa}
\def\l{\lambda}
\def\m{\mu}
\def\n{\nu}

\def\q{\theta}

\def\s{\sigma}

\def\x{\xi}

\def\F{\Phi}

\def\L{\Lambda}
\def\O{\Omega}

\def\S{\Sigma}
\def\U{\Upsilon}

\def\rd{{\rm d}}
\def\ri{{\rm i}}

\newcommand{\ve}{\varepsilon}                            

\newcommand{\pa}{\partial}                           
\newcommand{\hf}{\frac12}

\newcommand{\be}{\begin{equation}}
\newcommand{\ee}{\end{equation}}
\newcommand{\bea}{\begin{eqnarray}}
\newcommand{\eea}{\end{eqnarray}}
\newcommand{\non}{\nonumber}
\newcommand{\1}{{\underline{1}}}
\newcommand{\2}{{\underline{2}}}

\newcommand{\bm}[1]{\mbox{\boldmath$#1$}}

\def\double #1{#1{\hbox{\kern-2pt $#1$}}}

\newcommand{\hm}{{\hat{m}}}

\newcommand{\ha}{{\hat{a}}}
\newcommand{\hb}{{\hat{b}}}
\newcommand{\hc}{{\hat{c}}}
\newcommand{\hd}{{\hat{d}}}
\newcommand{\he}{{\hat{e}}}

\newcommand{\hal}{{\hat{\a}}}
\newcommand{\hbe}{{\hat{\b}}}
\newcommand{\hga}{{\hat{\g}}}
\newcommand{\hde}{{\hat{\d}}}

\begin{document}

\begin{titlepage}

\begin{flushright}
April, 2008\\
\end{flushright}
\vspace{5mm}

\begin{center}
{\Large \bf  Wandering in five-dimensional curved superspace}\footnote{Contribution to the proceedings of the 3$^{\rm rd}$ Workshop of the RTN project ``Constituents, Fundamental Forces and Symmetries of the Universe,'' Valencia, October 1-5, 2007.}\\ 
\end{center}

\begin{center}

{\large  
Sergei M. Kuzenko\footnote{{kuzenko@cyllene.uwa.edu.au}}
and 
Gabriele Tartaglino-Mazzucchelli\footnote{gtm@cyllene.uwa.edu.au}
} \\
\vspace{5mm}

\footnotesize{
{\it School of Physics M013, The University of Western Australia\\
35 Stirling Highway, Crawley W.A. 6009, Australia}}  
~\\

\vspace{2mm}

\end{center}
\vspace{5mm}

\begin{abstract}
\baselineskip=14pt
This is a brief review of the superspace formulation for 
five-dimensional  $\cN=1$ matter-coupled supergravity recently developed
by the authors.
\end{abstract}
\vspace{1cm}

\vfill
\end{titlepage}

\newpage
\renewcommand{\thefootnote}{\arabic{footnote}}
\setcounter{footnote}{0}

\section{Introduction}

Historically, the first attempt to formulate five-dimensional $\cN=1$ 
(often called $\cN=2$) supergravity 
in an off-shell superspace setting was made in \cite{BK} shortly 
before its on-shell component formulation was  given \cite{Cremmer,CN}.
Inspired by \cite{Cremmer}, 
Howe \cite{Howe5Dsugra} (see also \cite{HL})
proposed a superspace formulation 
for the minimal multiplet of 5D $\cN=1$ supergravity (``minimal'' in the sense 
of superconformal tensor calculus). After Howe's work \cite{Howe5Dsugra}, 
5D $\cN=1 $ curved superspace has been abandoned for 25 years.
General matter couplings in 5D $\cN=1$ supergravity have been constructed
within on-shell components approaches \cite{GST,GZ,CD}
and within the superconformal tensor calculus \cite{Ohashi,Bergshoeff}.

In 2007, we began the program of developing a superspace formulation for 
5D $\cN=1$ matter-coupled supergravity. We first elaborated supersymmetric field 
theory in 5D $\cN=1$ anti-de Sitter superspace which 
is a maximally symmetric curved background \cite{KT-M}.
This was followed  by a fully-fledged supergravity formalism developed in a 
series of papers \cite{KT-Msugra1,KT-Msugra2,KT-Msugra3}.
In these publications, we not only reproduced the main results of 
the superconformal tensor approach \cite{Ohashi,Bergshoeff}, 
but also proposed new off-shell supermultiplets and more general
supergravity-matter systems. The present note is a brief review of our construction.

Looking back at the 25 year history of   5D $\cN=1 $ curved superspace, 
one can notice a striking historical curiosity.  
In 1982, Howe had the right superspace setting for pure supergravity -- 
the minimal multiplet  \cite{Howe5Dsugra}, 
which was the starting point of our approach  \cite{KT-Msugra1,KT-Msugra2}.
The same multiplet  also occurs within the superconformal tensor calculus \cite{Ohashi,Bergshoeff}
by coupling the Weyl multiplet to an Abelian vector multiplet  and then gauge 
fixing some local symmetries (the vector multiplet is one of two compensators
required to describe  Poincar\'e supergravity).
 So why didn't Howe make use of his formulation 
to construct Poincar\'e supergravity and its matter couplings?
A partial answer is quite simple. Even in rigid supersymmetry 
with eight supercharges in diverse dimensions, 
 adequate approaches to generate off-shell supermultiplets
and supersymmetric actions appeared only in 1984.  
They go by the names  {\it harmonic superspace}
\cite{GIKOS,GIOS} and {\it projective superspace} \cite{KLR,LR}.

This note is organized as follows. In section 2 we review, following  \cite{KT-Msugra3},
the superspace formulation for the Weyl multiplet of conformal supergravity. 
Covariant projective supermultiplets and the supersymmetric action principle
are introduced in section 3. 
The same section also contains a few examples of interesting dynamical systems.

\section{5D conformal supergravity in superspace}
\label{Weyl}

We start by describing the superspace formulation 
for 5D conformal supergravity  \cite{KT-Msugra3}.
Let $z^{\hat{M}}=(x^{\hm},\q^{\hat{\mu}}_i)$ 
be local bosonic ($x$) and fermionic  ($\q$) coordinates 
parametrizing a curved five-dimensional $\cN=1$  superspace
$\cM^{5|8}$ ($\hm=0,1, \cdots,4$, $\hat{\mu}=1,\cdots,4$, and  $i=\1,\2$).
The Grassmann variables $\q^{\hat{\mu}}_i$ obey the  5D pseudo-Majorana 
reality condition
$\overline{\q^{\hat{\mu}}_i }= \q_{\hat{\mu}}^i =\ve_{\hat{\m} \hat{\n}}\ve^{ij}\q^{\hat{\nu}}_j  $.
The tangent-space group
is chosen to be  ${\rm SO}(4,1)\times {\rm SU}(2)$,
and the superspace  covariant derivatives 
$\cD_{\hat{A}} =(\cD_{\hat{a}}, \cD_{\hat{\a}}^i) $
have the form 
\bea
\cD_{\hat{A}}=
E_{\hat{A}} + \hf\O_{\hat{A}}{}^{\hb\hc}(z)M_{\hb\hc} + \F_{\hat{A}}{}^{kl}(z)J_{kl}
~.
\label{CovDev}
\eea
Here ${E}_{\hat{A}}= {E}_{\hat{A}}{}^{\hat{M}}(z) \pa_{\hat{M}}$ is the supervielbein, 
with $\pa_{\hat{M}}={\pa/\pa z^{\hat{M}}}$;
 $M_{\hb\hc}$ and  $\O_{\hat{A}}{}^{\hb\hc}$ are the Lorentz  generators and 
 connection respectively (both antisymmetric in $\hb$, $\hc$); 
 $J_{kl}$ and  $\Phi_{\hat{A}}{}^{kl}$ are respectively the SU(2) generator and connection
 (symmetric in $k$, $l$). 
The generators of ${\rm SO}(4,1)\times {\rm SU}(2)$
act on the covariant derivatives as follows:\footnote{The operation of
(anti)symmetrization of $n$ indices 
is defined to involve a factor $(n!)^{-1}$.}
\bea
{[}M_{\hal\hbe},\cD_{\hga}^k{]}=\ve_{\hga(\hal}\cD^k_{\hbe)}~,\qquad
{[}M_{\ha\hb},\cD_{\hc}{]}=2\eta_{\hc[\ha}\cD_{\hb]}~,\qquad
{[}J^{kl},\cD_{\hal}^i{]}= \ve^{i(k} \cD^{l)}_{\hat \a}~,
\label{generators}
\eea
where $J^{kl} =\ve^{ki}\ve^{lj} J_{ij}$ and $M_{\hal\hbe}=M_{\hbe\hal}
=(\S^{\ha\hb})_{\hal\hbe}M_{\ha\hb}$ and $(\S^{\ha\hb})_{\hal}{}^{\hbe}$
are the spinor Lorentz generators, $\S^{\ha\hb}=-\frac{1}{4} [\G^\ha , \G^\hb ]$,
with $\G^\ha $ the 5D Dirac matrices
 (see the appendix in \cite{KT-Msugra2} for  our notation 
and conventions).

The supergravity gauge group is generated by local transformations
of the form 
\be
\d_K \cD_{\hat{A}} =[ K, \cD_{\hat{A}} ]~,\quad
\d_K U = {K  }\, U~,
\qquad K = K^{\hat{C}}(z) \cD_{\hat{C}} +\hf K^{\hat c \hat d}(z) M_{\hat c \hat d}
+K^{kl}(z)J_{kl}
~,
\label{tau}
\ee
with all the gauge parameters 
obeying natural reality and symmetry conditions, and otherwise  arbitrary. 
In (\ref{tau}) we have also included the transformation rule for a tensor superfield $U(z)$, 
with its indices suppressed.

The covariant derivatives obey (anti)commutation relations of the general form 
\bea
{[}\cD_{\hat{A}},\cD_{\hat{B}}\}=T_{\hat{A}\hat{B}}{}^{\hat{C}}\cD_{\hat{C}}
+\hf R_{\hat{A}\hat{B}}{}^{\hat{c}\hat{d}}M_{\hat{c}\hat{d}}
+R_{\hat{A}\hat{B}}{}^{kl}J_{kl}
~,
\label{algebra}
\eea
where $T_{\hat{A}\hat{B}}{}^{\hat{C}}$ is the torsion, and 
$R_{\hat{A}\hat{B}}{}^{\hat{c}\hat{d}}$
and $R_{\hat{A}\hat{B}}{}^{kl}$ are
the  SO(4,1)  and SU(2) curvature tensors, respectively. 

To describe the  Weyl multiplet of conformal supergravity  \cite{Ohashi,Bergshoeff}, 
the torsion has to be constrained as
\cite{KT-Msugra3}:
\bea
T_{\hal}^i{}_{\hbe}^j{}^{\hc}~=~-2\ri\ve^{ij}(\Gamma^{\hc})_{\hal\hbe},  \qquad
T_{\hal}^i{}_{\hbe}^j{}^{\hga}_k~=~
T^i_{\hal}{}_{\hb}{}^{\hc}~
=~0,  \qquad  
T_{\ha\hb}{}^{\hc}~=~T_{\ha\hbe (j}{}^{\hbe}{}_{k)}~=~0 ~.
\label{constr}
\eea
${}$With these constraints, it can be shown that the torsion and curvature tensors 
are expressed in terms of four dimension-1 tensor superfields $S^{ij}$,  $C_\ha{}^{ij}$, $X_{\ha\hb}$, 
and $N_{\ha\hb}$, and their covariant derivatives. The superfields $S^{ij}$,  $C_\ha{}^{ij}$ 
are symmetric in $i,\,j$, while  $X_{\ha\hb}$, $N_{\ha\hb}$ are 
antisymmetric in $\ha\,,\hb$.
All these tensors are real $\overline{S^{ij}} =S_{ij}~, \overline{C_\ha{}^{ij} }=C_{\ha ij }~,
\overline{X_{\ha\hb}} =X_{\ha\hb}~,\overline{N_{\ha\hb}} =N_{\ha\hb}$.

The covariant derivatives obey the (anti)commutation relations \cite{KT-Msugra3}:
\begin{subequations}
\bea
\big\{ \cD_{\hal}^i , \cD_{\hbe}^j \big\} &=&-2 \ri \,\ve^{ij}\cD_{\hal\hbe}
-\ri \,\ve_{\hal\hbe}\ve^{ij}X^{\hc\hd}M_{\hc\hd}
+{\ri\over 4} \ve^{ij}\ve^{\ha\hb\hc\hd\he}(\G_\ha)_{\hal\hbe}N_{\hb\hc}M_{\hd\he}
\non\\
&&
-{\ri\over 2}\ve^{\ha\hb\hc\hd\he}(\S_{\ha\hb})_{\hal\hbe}C_{\hc}{}^{ij}M_{\hd\he}
+4\ri \,S^{ij}M_{\hal\hbe}
+3\ri \, \ve_{\hal\hbe}\ve^{ij}S^{kl}J_{kl}
\non\\
&&
-\ri \, \ve^{ij}C_{\hal\hbe}{}^{kl}J_{kl}
-4\ri\Big(X_{\hal\hbe}+N_{\hal\hbe}\Big)J^{ij}
~,
\label{covDev2spinor-} \\
{[}\cD_\ha,\cD_{\hbe}^j{]}&=&
{1\over 2} \Big(
(\Gamma_{\hat{a}})_{\hbe}{}^{\hga}S^j{}_k
- X_{\ha\hb}(\Gamma^{\hat{b}})_{\hbe}{}^{\hga} \d^j_k
-{1\over 4}\,\ve_{\ha\hb\hc\hd\he}N^{\hd\he}(\Sigma^{\hb\hc})_{\hbe}{}^{\hga}
\d^j_k
+ (\S_\ha{}^{\hb})_{\hbe}{}^{\hga}C_\hb{}^j{}_k
\Big)
\cD_{\hga}^k~~~~~~
\non\\
&&
\qquad \qquad 
~+~\mbox{curvature terms}~.
\label{covDev2spinor-2}
\eea
\end{subequations}
The dimension-1 components of the torsion, 
 $S^{ij}$, $X_{\ha\hb}$, 
$N_{\ha\hb}$ and $C_\ha{}^{ij}$, 
obey some differential constraints 
implied by the Bianchi identities \cite{KT-Msugra3}. 

The fact that the supergeometry introduced corresponds to 5D conformal supergravity, 
manifests itself in the invariance of  
the constraints (\ref{constr})  under infinitesimal 
super-Weyl transformations of the form\footnote{The finite form for the super-Weyl transformations 
has been given in \cite{KT-M-confFlat}.}
\begin{subequations}
\bea
\d_\s \cD_\hal^i&=&\s\cD_\hal^i+4(\cD^{\hga i}\s)M_{\hga\hal}-6(\cD_{\hal k}\s)J^{ki}~,
\label{sW1} 
\\
\d_\s \cD_\ha &=&
2\s\cD_\ha
+\ri(\G_\ha)^{\hga\hde}(\cD_{\hga}^{k}\s)\cD_{\hde k}
-2(\cD^\hb\s)M_{\ha\hb}
+{\ri\over 4}(\G_\ha)^{\hga\hde}(\cD_\hga^{(k}\cD_{\hde}^{l)}\s)J_{kl}
\label{sW2}
~,
\eea
\end{subequations}
where the scalar superfield $\s$ is real and unconstrained.
The components of the dimension-1 torsion can be seen to transform
as follows:
\begin{subequations}
\bea
\label{s-Weyl-Sij}
\d_\s S^{ij}&=&2\s S^{ij}
+{\ri\over 2}\,\cD^{\hal (i}\cD_{\hal}^{ j)}\s~,
\qquad \qquad  \quad
\d_\s C_{\ha}{}^{ij}=2\s C_{\ha}{}^{ij}
+{\ri}\,( \G_\ha)^{\hga\hde} \cD_{\hga}^{(i}\cD_{\hde}^{j)}\s~,
\\
\d_\s X_{\ha\hb}&=&2\s X_{\ha\hb}
-{\ri\over 2}\, (\S_{\ha\hb})^{\hal\hbe}\cD_\hal^k\cD_{\hbe k}\s~,
\qquad 
\d_\s N_{\ha\hb} = 2\s N_{\ha\hb}
-\ri \,(\S_{\ha\hb})^{\hal\hbe} \cD_{\hal}^{k}\cD_{\hbe k}\s~.
\label{s-Wey-N}
\eea
\end{subequations}
It follows from here that $W_{\ha \hb} := X_{\ha\hb} -\hf  N_{\ha\hb}$
transforms homogeneously, 
\be
\d_\s W_{\ha \hb} = 2\s W_{\ha \hb} ~.
\ee
Therefore, $W_{\ha\hb}$
is a superspace generalization of the Weyl tensor. 

It turns out that the super-Weyl transformations can be used to gauge away
the superfield $C_{\ha}{}^{ij}$.  Imposing 
the super-Weyl gauge condition
\bea
C_{\ha}{}^{ij}=0~,
\label{C=0}
\eea
is equivalent to extending the set of constraints  (\ref{constr}) by an additional dimension-1
constraint which is $T_\ha{}_{(\hbe}^{(j}{}_{\hga)}^{k)}=0$ \cite{KT-Msugra3}.
The resulting superspace geometry provides an alternative  
description of the Weyl multiplet. 
Because of (\ref{C=0}), the full set of constraints is now 
invariant under the super-Weyl transformations (\ref{sW1})--(\ref{sW2}) generated by  
a constrained parameter $\s$.
The corresponding constraint is
\bea
\cD_{\hal}^{(i}\cD_{\hbe}^{j)}\s
-{1\over 4}\ve_{\hal\hbe}\cD^{\hga(i}\cD_{\hga}^{j)} \s =0 ~.
\label{s-C=0}
\eea
Another consequence of (\ref{C=0}) in conjunction with  the Bianchi identities is that 
 $S^{ij}$ satisfies the equation
\bea
\cD_\hga^{(i}S^{jk)}=0~.
\label{3/2Dev+S++}
\eea
If not specifically mentioned, eq. (\ref{C=0}) will be assumed in what follows.

The Weyl multiplet can naturally be coupled to a non-Abelian vector multiplet.
This is achieved by introducing gauge-covariant derivatives 
${\bm \cD}_{\hat A} = \cD_{\hat A} +\cV_{\hat A}(z)$, with $\cV_{\hat A}$ 
a gauge connection taking its values in the Lie algebra of the gauge group.
Then the algebra (\ref{algebra}) turns into
\bea
{[}{\bm \cD}_{\hat{A}},{\bm \cD}_{\hat{B}}\}=T_{\hat{A}\hat{B}}{}^{\hat{C}}{\bm \cD}_{\hat{C}}
+\hf R_{\hat{A}\hat{B}}{}^{\hat{c}\hat{d}}M_{\hat{c}\hat{d}}
+R_{\hat{A}\hat{B}}{}^{kl}J_{kl} + \cF_{\hat{A}\hat{B}}~.
\label{algebra2}
\eea
An irreducible off-shell vector multiplet emerges if 
$\cF_{\hat{A}\hat{B}}$ is constrained as 
$\cF_{\hal}^i{}_{\hbe}^j \propto \ve^{ij} \ve_{\hat \a \hat \b} \cW$
(compare with \cite{HL}). 
The field strength $\cW$
possesses the 
super-Weyl transformation $\d_\s \cW=2\s \cW$ and obeys the following Bianchi identity:
\bea
{\bm \cD}_{\hal}^{(i}{\bm \cD}_{\hbe}^{j)}\cW
-{1\over 4}\ve_{\hal\hbe}{\bm \cD}^{\hga(i}{\bm \cD}_{\hga}^{j)} \cW = 0~.
\label{W-BI}
\eea
Associated with the vector multiplet  is the composite 
superfield  \cite{KT-Msugra3} 
\bea
\cG^{ij}:= {\rm tr} \,\Big\{ 
\ri\,{\bm \cD}^{\hal (i}\cW {\bm \cD}_\hal^{j)} \cW
+{\ri\over 2} \cW {\bm \cD}^{ij} \cW-2S^{ij}\cW^2 \Big\}~,
\qquad {\bm \cD}^{ij}:= {\bm \cD}^{\hal (i} {\bm \cD}_\hal^{j)}  ~.
\label{Gij}
\eea
It is characterized by the following fundamental properties:
\bea
\cD^{(i}_\hal \cG^{jk)}=0~,\qquad
\d_\s \cG^{ij}&=&{6\s} \cG^{ij}~.
\label{G-anal}
\eea

Let $\cW= W \,{\bm Z}$, with ${\bm Z}$ the generator,
be the field strength of an Abelian vector multiplet.
Then, eq. (\ref{W-BI}) coincides in form with the constraint (\ref{s-C=0})
obeyed by the super-Weyl parameter. 
If the vector multiplet is characterized by $W(z) \neq 0$ everywhere in superspace,
super-Weyl transformations can be used  to impose the gauge $W=1$.
The resulting  geometry (\ref{algebra2})  describes the minimal 
multiplet of 5D supergravity \cite{Howe5Dsugra}.

\section{Kinematics and dynamics in curved  projective superspace}
\label{projective}

We have reviewed the geometric description of 5D conformal supergravity 
in superspace. Let us now turn to a brief discussion of a large family of off-shell
supermultiplets coupled to conformal supergravity, which can be used 
to describe supersymmetric matter. They were introduced in   \cite{KT-Msugra3}
under the name  {\it covariant projective supermultiplets}. These supermultiplets
are a curved-superspace extension of the 5D superconfomal projective multiplets
\cite{K}. The latter are ordinary projective supermultiplets \cite{LR} 
with respect to the super-Poincar\'e subgroup of the 5D superconformal 
group.

It is useful to introduce auxiliary 
isotwistor coordinates
$u^{+}_i \in  {\mathbb C}^2 \setminus  \{0\}$
in addition to the superspace coordinates $z^{\hat{M}}=(x^{\hm},\q^{\hat{\mu}}_i)$.
All the coordinates $u^{+}_i$ and $z^{\hat{M}}$ are defined to be inert 
under the tangent-space group. In particular,
the variables $u^{+}_i$ do not transform under
the local SU(2)  group, and hence they are covariantly constant,
$\cD_{\hat A} u^+_j=0$. 
It follows from  (\ref{covDev2spinor-}) that the operators 
$\cD^+_{\hat \a}:=u^+_i\,\cD^i_{\hat \a} $  obey the following algebra
(the constraint  (\ref{C=0}) is not assumed from here until eq. (\ref{super-Weyl-Qn}) including):
\bea
\{  \cD^+_{\hat \a} , \cD^+_{\hat \b} \}
=-4{\rm i}\, \Big(X_{\hal \hbe}+N_{\hal \hbe}\Big)\,J^{++}
+4{\rm i} \, S^{++}M_{\hal \hbe}
-{\ri\over 2}\ve^{\ha\hb\hc\hd\he}(\S_{\ha\hb})_{\hal\hbe}C_\hc{}^{++}M_{\hd\he}~,
\label{analyt}
\eea
where 
$J^{++}:=u^+_i u^+_j \,J^{ij}$ and 
$S^{++}:=u^+_i u^+_j \,S^{ij}$.

A covariant  projective supermultiplet of weight $n$,
$Q^{(n)}(z,u^+)$, is defined to be a scalar superfield that 
lives on  $\cM^{5|8}$, 
is holomorphic with respect to 
the isotwistor variables $u^{+}_i $ on an open domain of 
${\mathbb C}^2 \setminus  \{0\}$, 
and is characterized by the following conditions:\\
(i) it obeys the covariant analyticity constraint
\be
\cD^+_{\hat \a} Q^{(n)}  =0~;
\label{ana}
\ee  
(ii) it is  a homogeneous function of $u^+$ 
of degree $n$, that is,  
\be
Q^{(n)}(z,c\,u^+)\,=\,c^n\,Q^{(n)}(z,u^+)~, \qquad c\in
{\mathbb C}\setminus  \{0\}
~;
\label{weight}
\ee
(iii) infinitesimal gauge transformations (\ref{tau}) act on $Q^{(n)}$ 
as follows:
\bea
\d_K Q^{(n)} 
&=& \Big( K^{\hat{C}} \cD_{\hat{C}} + K^{ij} J_{ij} \Big)Q^{(n)} ~,  
\non \\ 
K^{ij} J_{ij}  Q^{(n)}&=& -\frac{1}{(u^+u^-)} \Big(K^{++} D^{--} 
-n \, K^{+-}\Big) Q^{(n)} ~, \qquad 
K^{\pm \pm } =K^{ij}\, u^{\pm}_i u^{\pm}_j ~,
\label{harmult1}   
\eea 
where
$D^{--}=u^{-i}{\partial}/{\partial u^{+ i}}$. 
The right-hand side in  (\ref{harmult1}) involves an additional isotwistor,  $u^-_i$ 
which is subject to the condition $(u^+u^-) = u^{+i}u^-_i \neq 0$, 
and is otherwise arbitrary.
By construction, $Q^{(n)}$ is independent of $u^-$, 
i.e. $\pa  Q^{(n)} / \pa u^{-i} =0$.
One can see that $\d Q^{(n)} $ 
is also independent of the isotwistor $u^-$, 
that is $\pa (\d Q^{(n)})/\pa u^{-i} =0$,
due to (\ref{weight}). 
It follows from (\ref{harmult1}) that $J^{++} \,Q^{(n)}\equiv0$ 
which is the integrability condition 
for the constraint (\ref{ana}).
It is important to note that, 
because of (ii), 
the isotwistor $u^+_i$ plays the role of homogeneous global coordinates for 
$\mathbb{C}P^{1}$ and the covariant projective multiplets 
live in curved projective superspace
$\cM^{5|8}\times\mathbb{C}P^{1}$.

In the case of conformal supergravity, we have to address the issue of how  
covariant projective multiplets may consistently 
vary under the super-Weyl transformations. 
If a weight-$n$ projective superfield $Q^{(n)}$ is chosen to
transforms homogeneously, $\d_\s Q^{(n)}\propto  \s Q^{(n)}$,
then its transformation law turns out to be  uniquely fixed 
by the constraint  (\ref{ana}) to be
\bea
\d_\s Q^{(n)}=3n \,\s Q^{(n)}~.
\label{super-Weyl-Qn}
\eea
Without the assumption of homogeneity, 
it is easy to construct examples of covariant projective multiplets which do not respect 
(\ref{super-Weyl-Qn}). 
The superfield $S^{++}$ is a particularly important example.
Due to eq. (\ref{3/2Dev+S++}) (from here on we only consider the geometry with $C_\ha{}^{ij}=0$),
$S^{++}$ is a projective superfield of weight two, 
$\cD^+_{\hat \a} S^{++}  =0$.
In accordance with (\ref{s-Weyl-Sij}), its super-Weyl transformation is inhomogeneous
\be
\d_\s S^{++}=2\s S^{++}
+{\ri\over 2}\,(\cD^+)^2\s~, \qquad 
(\cD^+)^2:=\cD^{+\hal}\cD^+_{\hal}~.
\label{s-Weyl-S++}
\ee

Another important example of weight-two projective multiplet is given by
$
\cG^{++} := \cG^{ij}u^+_iu^+_j
$
with  $\cG^{ij}$ the descendant 
associated with the Yang-Mills
field strength $\cW$ defined in (\ref{Gij}).
It satisfies the constraint $\cD^+_{\hat \a} \cG^{++}  =0$,
and possesses
the super-Weyl transformation law 
$\d_\s \cG^{++}=6 \s \cG^{++}$
\cite{KT-Msugra3}.

If  $Q^{(n)}(u^+)$ is a covariant projective multiplet, 
its complex conjugate $\bar{Q}^{(n)}(\overline{u^+}) $
is no longer of the same type.
However, 
one can introduce a generalized
{\it smile}-conjugation, $Q^{(n)} \to \widetilde{Q}^{(n)}$, 
\be
\widetilde{Q}^{(n)} (u^+)\equiv \bar{Q}^{(n)}\big(
\overline{u^+} \to 
\widetilde{u}^+\big)~, 
\qquad \widetilde{u}^+ = {\rm i}\, \s_2\, u^+~, 
\ee
which acts on the space of covariant projective weight-$n$ multiplets, 
since $\widetilde{ {\cD^+_{\hat \a} Q^{(n)}} }=(-1)^{\e(Q^{(n)})}\, \cD^{+\hat \a}
 \widetilde{Q}{}^{(n)}$. 
One can see that
$\widetilde{\widetilde{Q}}{}^{(n)}=(-1)^nQ^{(n)}$,
and therefore real supermultiplets can be defined for 
$n$  even.

To define a locally supersymmetric and super-Weyl invariant action, 
one needs two prerequisites   \cite{KT-Msugra3}: (i) a Lagrangian 
 $\cL^{++} (z,u^+)$ which is a real projective multiplet of weight two
and which possesses the super-Weyl transformation $\d_\s \cL^{++}=6\s \cL^{++}$; 
(ii)  an Abelian vector multiplet  with its field strength $W(z)$ 
non-vanishing everywhere. The action is:
\bea
S(\cL^{++})&=&
\frac{2}{3\pi} \oint (u^+ \rd u^{+})
\int \rd^5 x \,{\rm d}^8\q\,E\, \frac{\cL^{++}\,W^4}{(G^{++})^2}~, 
\qquad E^{-1}={\rm Ber}\, (E_{\hat A}{}^{\hat M})~.
\label{InvarAc}
\eea
Here $G^{++} := G^{ij}u^+_iu^+_j$, where $G^{ij}$  is the descendant (\ref{Gij})
associated with $W$.
Note that $S(\cL^{++})$ is    invariant under arbitrary re-scalings
$u_i^+(t)  \to c(t) \,u^+_i(t) $, 
$\forall c(t) \in {\mathbb C}\setminus  \{0\}$, 
where $t$ denotes the evolution parameter 
along the integration contour.
The action can be shown to be invariant  
under   supergravity gauge transformations
(\ref{tau}) and (\ref{harmult1}),  see \cite{KT-Msugra3,KT-Msugra2}.
To see that $S(\cL^{++})$ is invariant  under super-Weyl transformations, 
one has only to note that $\d_\s E=-2\s E$ and 
make use of the transformation rules
$\d_\s \cL^{++}=6\s \cL^{++}$, $\d_\s W=2\s W$ and $\d_\s G^{++}=6\s G^{++}$.

The crucial property of $S(\cL^{++})$ 
 is that it is  independent of the concrete choice  of $W$,
provided $\cL^{++}$ is independent of such a vector multiplet.
Another important feature of the action introduced is
that (\ref{InvarAc}) provides a natural extension of the 
action principle in flat projective superspace \cite{KLR,K}.

Since the action  (\ref{InvarAc}) is super-Weyl invariant, 
one can choose 
the super-Weyl gauge 
$W=1$.
Then, the action functional (\ref{InvarAc}) takes the form 
given in \cite{KT-Msugra2} in the case of the 5D minimal multiplet.

Now we are in a position to  give some interesting examples of supergravity-matters systems.
Let  ${\mathbb V}(z,u^+)$ denote the tropical prepotential\footnote{See 
\cite{KT-Msugra1}
for the definition of covariant arctic and tropical multiplets.}
for the Abelian vector multiplet $W$ appearing in the action  (\ref{InvarAc}). 
The prepotential 
is a real weight-zero projective multiplet 
possessing the gauge invariance 
\be
\d {\mathbb V} =  \l +\tilde{\l}~, 
\ee
with $\l$ a weight-zero arctic multiplet.  
A hypermultiplet can be described 
by an arctic weight-one multiplet 
$\U^{+ } (z,u^+) $ and its smile-conjugate 
$ \widetilde{\U}^{+}$.  
Consider a  gauge invariant Lagrangian of the form 
(with the gauge transformation of $\U^+$ being $\d\U^+=-{\xi}\l\U^+$)
\be
\cL^{++} = \frac{1}{k^2} {\mathbb V} \,G^{++} 
-  \widetilde{\U}^{+} {\rm e}^{\x {\mathbb V} }\, \U^{+ }~,
\ee
with $\k$ the gravitational coupling  constant, and $\x$ a cosmological constant. 
It describes Poincar\'e supergravity if $\x=0$, and 
pure gauge supergravity with $\x\neq0$.

The dynamics of the Yang-Mills 
supermultiplet can be described by the Lagrangian
$
\cL^{++}_{\rm YM} = g^{-2}\, {\mathbb V} \, \cG^{++} ,
$
with $g$  the coupling constant (compare with the rigid
supersymmetric case \cite{KL}).

A system of  arctic weight-one multiplets 
$\U^{+ } (z,u^+) $ and their smile-conjugates
$ \widetilde{\U}^{+}$ can be   described by the Lagrangian
\bea
\cL^{++} = {\rm i} \, K(\U^+, \widetilde{\U}^+)~,
\label{conformal-sm}
\eea
with $K(\F^I, {\bar \F}^{\bar J}) $ a real analytic function
of $n$ complex variables $\F^I$, where $I=1,\dots, n$.
${}$For $\cL^{++}$ to be a weight-two real projective superfield, 
it is sufficient to  require 
\bea
 \F^I \frac{\pa}{\pa \F^I} K(\F, \bar \F) =  K( \F,   \bar \F)~.
 \label{Kkahler2}
 \eea
This is a curved superspace generalization of the general model 
for  superconformal polar multiplets
\cite{K} (see also \cite{KT-M}).

Given a system of interacting arctic weight-zero multiplets 
${\bf \U}  $ and their smile-conjugates
$ \widetilde{ \bf{\U}}$, their coupling to supergravity can be described by the Lagrangian 
\bea
\cL^{++} = G^{++}\,
{\bf K}({\bf \U}, \widetilde{\bf \U})~,
\eea
with ${\bf K}(\F^I, {\bar \F}^{\bar J}) $ a real function 
which is not required to obey any 
homogeneity condition. 
The corresponding action is invariant under K\"ahler transformations of the form
\be
{\bf K}({\bf \U}, \widetilde{\bf \U})~\to ~{\bf K}({\bf \U}, \widetilde{\bf \U})
+{\bf \L}({\bf \U}) +{\bar {\bf \L}} (\widetilde{\bf \U} )~,
\ee
with ${\bf \L}(\F^I)$ a holomorphic function.\\

\noindent
{\bf Acknowledgements:}\\
G.T.-M. would like to thank the organizers of the 3$^{\rm rd}$ RTN Workshop ``Constituents, Fundamental Forces and Symmetries of the Universe'' for the opportunity to report
on preliminary results of the research reviewed in this contribution.
This work is supported by the Australian Research Council.


\small{

}

\end{document}